\documentclass[a4paper,12pt]{article}

\usepackage{amsmath}
\usepackage{graphicx}
\usepackage{bm}
\usepackage[margin=15pt,font=small,labelfont=bf]{caption}
\newcommand{\mbf}{\boldsymbol}
\begin{document}

\title{Efficient extraction of a collimated ultra-cold neutron beam using diffusive channels}
\author{P. Schmidt-Wellenburg\thanks{\textit{Corresponding author: Tel.: +33
4 76 20 70 27; fax: +33 4 76 20 77 77. E-mail address: schmidt-w@ill.fr}}~$%
^{1,2}$, J. Barnard$^1$, P. Geltenbort$^1$, V.V. Nesvizhevsky$^1$,\\
\medskip  C. Plonka$^1$, T. Soldner$^1$, O. Zimmer$^2$\\
\textit{\small 1) Institut Laue Langevin,}\\
\medskip \textit{\small 6, rue Jules Horowitz, BP-156, 38042 Grenoble
Cedex 9, France}\\
\textit{\small 2) Physik-Department E18,}\\
\textit{{\small Technische Universit\"{a}t M\"{u}nchen, 85748 Garching,
Germany}}\\
}
\maketitle

\abstract{We present a first experimental demonstration of a new method to extract a well-collimated beam of ultra-cold neutrons (UCN) from a storage vessel. Neutrons with too large divergence are not removed from
the beam by an absorbing collimation, but a diffuse or semidiffuse channel with high Fermi potential reflects them back into the vessel. This avoids unnecessary losses and keeps the storage time high, which may be beneficial when the vessel is part of a UCN source with long buildup time of a high UCN density. }\\

\noindent
\textbf{PACS: 28.20.-v, 28.20.Gd, 29.25.Dz, 28.41.Rc } \\

\setlength{\baselineskip}{2em}

Many next-generation ultra-cold neutron sources employ conversion of cold
neutrons in solid deuterium or superfluid helium down to the ultra-cold
regime $E_{\mathrm{\scriptscriptstyle UCN}}\leq 250$~neV\,(see, e.g., refs. \cite{Trinks/2000,Masuda/2002,Saunders/2004,Atchison/2005}
and the book \cite{Golub}). In a superthermal source UCN are accumulated in
a storage vessel to build up a high UCN density, from where they are
extracted continuously or periodically, depending on the specific
experiment. The saturated UCN density is given by

\begin{equation}
\rho _{\mathrm{\scriptscriptstyle UCN}}=P\tau _{\mathrm{tot}},
\end{equation}%
where $P$ is the production rate per unit volume and $\tau _{\mathrm{tot}}$ the storage lifetime of the vessel. Any aperture reduces the storage time $\tau _{\mathrm{tot}}$
and thus the maximum density, which in turn reduces the extracted flux. If
an experiment requires only a UCN beam with narrow divergence, extraction
without loosing unwanted neutrons would be advantageous. This can be
realised using a channel made of rough walls of high Fermi potential. Neutrons
scattered from the rough surface will start a diffuse motion inside the channel
and, for a sufficiently long channel, be scattered back into the primary
volume, unless they decay. For definition of a beam in one dimension, we
employ a channel made from two parallel surfaces, either both rough or one
rough the other specular (called diffuse and semidiffuse). A diffuse channel of height~$h$ and length~$l$ will transmit a beam with a divergence of $\mbf{p}_{\bot}\leq \mbf{p}_{||}\cdot h/l$. In a horizontal configuration of same height with a specular lower surface the angle of divergence is doubled, and neutrons with $E_{\bot}\leq mgh$ will also be transmitted.
Such a semidiffuse horizontal channel might be ideal for the spectrometer GRANIT \cite{Nesvizhevsky}, currently being developed to
measure transitions between gravitationally bound quantum states of UCN above a
mirror. Losses may appear inside the channel and due to outward diffusion of neutrons.

\begin{figure}[t]
\centering
\includegraphics{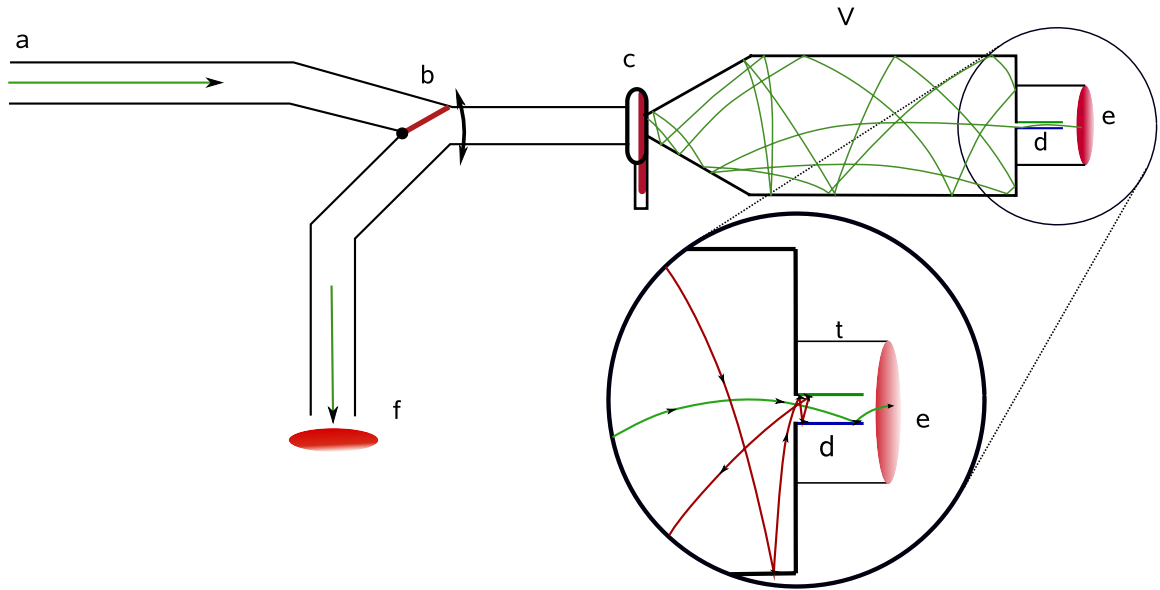} 
\caption{Experimental set-up (not to scale): UCN pass through the beam
guide~(a) and a switch~(b), which is used to either fill the storage
volume~(V) or empty it into the "storage" detector~(f). An UCN tight
shutter~(c) is used to lock off the storage volume. The extraction
channel~(d) is made of two rectangular pieces of sapphire (w:\,80~mm, l:\,30~mm, h:\,8~mm), polished on one and rough
on the other side. They are externally connected to the Fomblin coated
storage volume. The "transmission" detector~(e) is fixed to the volume via
an aluminium tube~(t) (\o {}\thinspace 90~mm, length: 80~mm) which
is covered with a PET foil to eliminate randomly scattered UCN.}
\label{fig:Setup}
\end{figure}

The storage time $\tau _{\mathrm{tot}}$ of UCN in a vessel with an
extraction channel can be described by 

\begin{equation}
1/\tau _{\mathrm{tot}}=1/\tau _{\mathrm{vol}}+1/\tau _{\mathrm{ch}},
\label{eqn:storage}
\end{equation}%
where $\tau _{\mathrm{vol}}$ is the storage time of the vessel without
extraction channel. The loss rate $1/\tau _{\mathrm{ch}}$ depends on the
probability $r$ that a UCN entering the channel becomes reflected back into
the vessel (in the following $r$ is simply called "reflectivity") and the
cross section $A_{\mathrm{ch}}$ of the channel,
 
\begin{equation}
1/\tau _{\mathrm{ch}}=\left( 1-r\right) \frac{\bar{v}_{\mathrm{\scriptscriptstyle UCN}}A_{ \mathrm{ch}}}{4V}. 
 \label{eqn:refl}
\end{equation}
$V$ is the storage volume and $\bar{v}_{\mathrm{\scriptscriptstyle UCN}}$ the mean UCN velocity. The reflectivity $r$ can be deduced experimentally from comparison of the semidiffuse~(or diffuse) configuration with the specular at same height. It can not be directly deduced from the transmission because of unknown losses due to wall absorption and up-scattering. 
A first experiment was carried out at the ultra-cold neutron installation
PF2 \cite{Steyerl/1986} of the Institut Laue Langevin, Grenoble. The set-up
is shown in Fig. \ref{fig:Setup}.

The channel was made from sapphire plates which have a Fermi potential larger than that of the Fomblin grease coating of the storage vessel. Each of the plates has a rough and a specular surface. Thus it is possible to measure with two rough~(double diffuse, dd) or two specular~(sp) surfaces, or with the semidiffuse (sd) configuration
with a rough upper and a specular~lower surface. Channel heights $100~%
\mathrm{\mu m}$, $500~\mathrm{\mu m}$ and $1000~\mathrm{\mu m}$, defined by
spacers, were used in the sd and sp configurations, the height of $1000~%
\mathrm{\mu m}$ also in the dd configuration. A position sensitive
CASCADE\,-\,U detector\,\cite{Schmidt/2003} detected the UCN transmitted through the channel. Its readout structure consists of 30 horizontal
stripes covering a height of $90~\mathrm{mm}$. The stripes were merged
to six larger zones with 30, 9, 9, 9, 9, and 24~mm height, respectively. The
channel setup could be manually changed by opening the vacuum on the side of the
transmission detector. The horizontal alignment was checked with a precision
spirit level. Thereafter the whole setup was evacuated below $5\times
10^{-4} $~mbar. 
\begin{table}[h]
\centering
\begin{tabular}{|r|r@{.}l@{~$\pm$~}r@{.}l|c|r@{.}l|}
\hline
channel height [$\mathrm{\mu m}$] & \multicolumn{4}{c|}{storage time $\tau_{%
\mathrm{tot}}$ [s]} & channel type & \multicolumn{2}{c|}{reflectivity} \\ 
\hline
100 & 110 & 2 & 2 & 2 & sd & 0 & $98^{+0.02}_{-0.16}$ \\ 
100 & 94 & 7 & 1 & 7 & sp & \multicolumn{2}{c|}{} \\ 
500 & 90 & 1 & 0 & 8 & sd & 0 & $39\pm0.06$ \\ 
500 & 80 & 5 & 1 & 6 & sp & \multicolumn{2}{c|}{} \\ 
1000 & 80 & 9 & 1 & 3 & sd & 0 & $35\pm0.05$ \\ 
1000 & 70 & 8 & 1 & 2 & sp & \multicolumn{2}{c|}{} \\ 
1000 & 92 & 4 & 1 & 8 & dd & 0 & $65\pm0.05$ \\ 
0 & 110 & 9 & 2 & 3 &  & \multicolumn{2}{c|}{} \\ \hline\hline
\end{tabular}%
\caption{Storage times of the Fomblin coated volume for different measured channel heights and types: semidiffuse~(sd), specular~(sp), and double diffuse~(dd). The reflectivity is given relative to the specular channel of same height.}
\label{tab:time}
\end{table}

For each configuration, we measured during an emptying time of $200$~s the
number of UCN remaining in the vessel, having stored them with the
shutter C closed for a holding time of 50, 100, 150, 200, 250, or 300~s. After correcting each data point for
losses during counting, the storage time was obtained from a fit of a single
exponential decay to the data. The results for the different configurations
are listed in Table \ref{tab:time}. The channel height $0\mathrm{\mu m}$
corresponds to the closed volume (the two plates are placed on top of each
other without spacer). This yields the unpertubed storage time $\tau _{%
\mathrm{vol}}$. From equations (\ref{eqn:storage}) and (\ref{eqn:refl}),
replacing $\tau _{\mathrm{ch}}$ by $\tau _{\mathrm{sp}}$ for the specular,
and by $\tau _{\mathrm{sd}}$ for the semidiffuse channel, the reflectivity
is given as 

\begin{equation}
r=1-\frac{\tau _{\mathrm{sp}}}{\tau _{\mathrm{sd}}},
\end{equation}%
where we have assumed $r=0$ for the specular channel. For all channel heights we observed
an increase of the storage lifetime of the semidiffuse configuration with respect
to the specular one. For a channel height of $100~\mathrm{\mu m}$ the reflectivity
amounts to $r\geq 80$~\%. This channel does not reduce the storage time of
our vessel any more. 
\begin{figure}[tbp]
\centering
\includegraphics[width=0.9\textwidth]{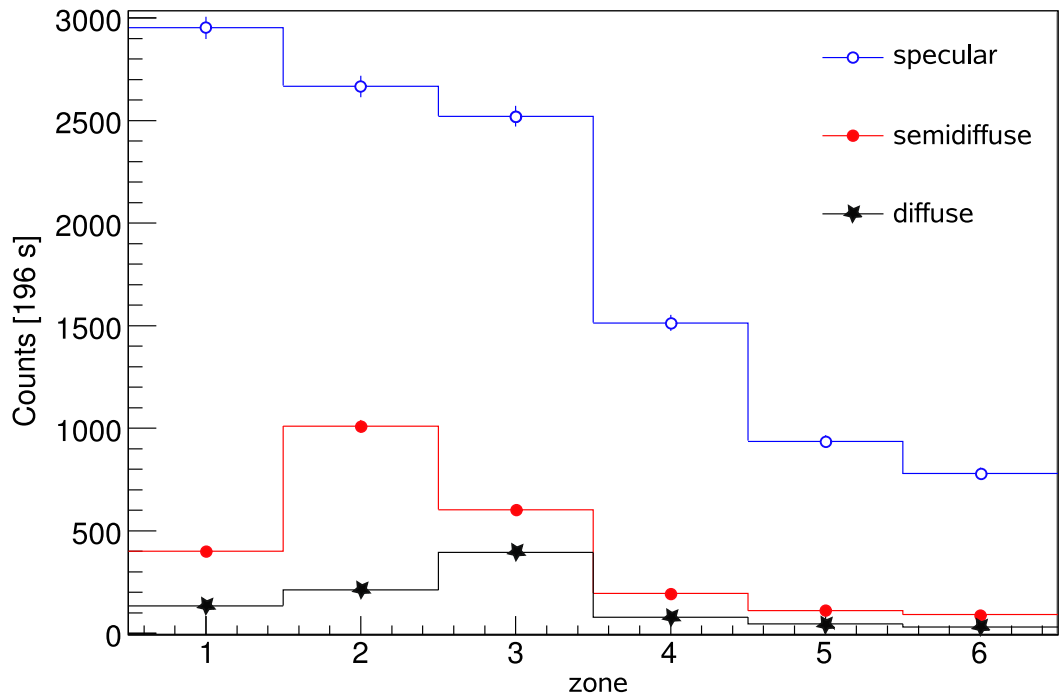} 
\caption{Vertical distribution of transmitted neutrons for the $1000~\mathrm{%
\protect\mu m}$ channel 50~mm behind channel exit. The channel exit window was in between detection
zone 2 and 3. Zone 1 is on the top, whereas zone 6 is on the bottom.
Furthermore, zone 1 and 6 had larger active areas\,(see text). The resolution is not sufficient to resolve the collimated beam.
}
\label{fig:trans}
\end{figure}

Figure \ref{fig:trans} shows the vertical distribution of transmitted UCN.
For the same channel height, the total transmission
decreases with addition of roughness to the surfaces. Furthermore, the
intensity distribution shows an attenuation of UCN with larger vertical
momentum if the surfaces are changed from specular to semidiffuse, and to
completely diffuse configuration. The observed spatial distribution
demonstrates the collimating properties of the semidiffuse and diffuse
channels.

In summary, we have shown that one may continuously extract, by a
semidiffuse or diffuse channel, a well-collimated beam of ultra-cold
neutrons from a vessel without significantly diminishing its storage time.
We measured UCN reflectivities from the channel opening better than$\ 80$ \%
in this first experimental test of a new method to be explored further in
future experiments.

\end{document}